\documentclass[showpacs,prb,aps,amsmath,amssymb,twocolumn]{revtex4-1}

\usepackage{subfigure}
\usepackage{amsmath}

\usepackage{subfigure}
\usepackage{amsmath}
\usepackage{graphicx}
\usepackage{rotating}

\newcommand{\beq}{\begin{eqnarray}}
\newcommand{\eeq}{\end{eqnarray}}

\newcommand{\br}{{\bf r}}

\newcommand{\beqa}{\begin{eqnarray}}
\newcommand{\eeqa}{\end{eqnarray}}

\newcommand {\be}{\begin{equation}}
\newcommand {\ee}{\end{equation}}
\newcommand {\bea}{\begin{eqnarray}}
\newcommand {\eea}{\end{eqnarray}}

\begin{document}

\title{Superconductivity in bundles of a mixture of doped carbon
nanotubes}
\author{Ilya Grigorenko}
\affiliation{Department of Physics and Materials Research Institute, Pennsylvania State University, University Park,  PA 16802, USA}
\author{Anvar Zakhidov}
\affiliation{UTD-NanoTech Institute, The University of Texas at Dallas, Richardson, Texas 75083-0688, USA}
\date{\today}

\begin{abstract}
 Using inhomogeneous Bogoliubov-de Gennes formalism we study superconducting properties of  bundles of single wall 
carbon nanotubes, consisting of a mixture of metallic and semiconducting nanotubes, having different 
critical transition temperatures. 
We investigate how the averaged superconducting order parameter and the critical transition temperature depend on the
 fraction of semiconducting carbon nanotubes in the bundle.
\end{abstract}
\maketitle
%\section{Introduction}

Single wall carbon nanotubes (SWCNT) represent a unique class of quasi-one dimensional nanoscale systems, 
exhibiting various interesting phenomena. Among other exciting features, 
it was demonstrated  that individual single wall carbon nanotubes may
have intrinsic superconducting properties \cite{4a_nanotube}. However, because of their extremely 
small diameter (just few nanometers), and thus strongly one dimensional character, the
superconducting order parameter may have significant ``phase slips'' due to thermal and quantum fluctuations, 
leading to a 
finite conductivity in the system below the critical temperature \cite{qf,thermal_slips}. Carbon nanotubes can 
form bundles and ropes 
\cite{rope},
 with tens and hundreds of individual SWCNTs in the bundle, coupled to each other by dispersive Van der Waals forces.
Such kind of system may exhibit reduced ``phase slips'' effects, due to three-dimensional  coupling of the nanotubes in the bundle 
and as a result, much stronger conductivity drop below the critical temperature.
The overall length of a SWCNT in the bundle also plays a significant role. 
For example, reducing the bundle's length to 300 nm destroys the superconductivity 
in the system due to increasingly high quantum fluctuations \cite{qf}. Generally speaking,  for nanoscale 
systems with the quantum level spacing approaching the superconducting gap energy $\Delta$, 
the superconductivity vanishes \cite{small_grains}.

It is expected that doping of SWCNTs in a bundle by, for example, boron, 
may significantly improve their superconducting properties.
It is believed that a proper level of doping may result in the Fermi level at a one dimensional singularity 
of the energy spectrum and thus in a higher density of states (DOS), that will lead to a higher 
critical temperature $T_c$. In particular, we suggest here that such kind of mechanism of doping enhanced 
$T_c$ may be much better pronounced in the case of semiconducting SWCNTs, 
which may have higher DOS due to lower in energy van Hove singularities. This is in contrast to  metallic SWCNTs, where 
 singularities in the DOS are much higher in energy, and start being filled much later during the doping process 
(according to the famous Kataura plot \cite{kataura}). 
Therefore, a bundle consisting of doped semiconducting nanotubes could be a much better superconductor, 
compared to a bundle made of metallic SWCNTs.
 
However,  synthesis of SWCNTs by all currently 
known methods results in a mixture of  semiconducting and metallic nanotubes. Since  
the nanotubes after the synthesis initially are not doped (or unintentionally slightly p-type doped, e.g. by oxygen of atmosphere),
 those are only metallic tubes, which may have superconducting transition, while semiconducting tubes will be ``diluting'' 
superconductivity in the bundle by the inverse proximity effect \cite{proximity}.   
Upon doping (i.e. by electrochemical charging), the 
semiconducting tubes can become superconducting with a higher  superconducting gap and
 thus a higher $T_c$  than in metallic nanotubes. 
%%%%%%%%%%%%%%%%%%%%%%%%%%%%%%%%%%%%%%%%%%%%%%%%%%%%%%%%%%%%%%%%%%%%%%%%%%%%%%%%%%
%%%%%%%%%%%%%%%%%%%%%%%%%%%%%%%%%%%%%%%%%%%%%%%%%%%%%%%%%%%%%%%%%%%%%%%%%%%%%%%%%%%%%%%%%%%
%%%%%%%%%%%%%%%%%%%%%%%%%%%%%%%%%%%%%%%%%%%%%%%%%%%%%%%%%%%%
%%%%%%%%%%%%%%%%%%%%%%%%%%%%%%%%%%%%%%%%%%%%%%
%%%%%%%%%%%%%%%%%%%%%%%%%%%%%%%
%%%%%%%%%%%%%%%%%
%%%%%%%

%\section{Model}

%As we discussed above, 
%we consider  two types of nanotubes in the bundle. 
%If the bundle $100\%$ consists of the first type of nanotubes (doped semiconducting), 
%we assume it has the order parameter $\Delta^1$ at temperature $T=0$ and corresponding critical temperature $T^1_c$,
% and if it solely consists of the second type (metallic)- 
%it has the order parameter  $\Delta^2<\Delta^1$ and the critical temperature $T^2_c<T^1_c$.
Therefore, one should be able to estimate spatially averaged order parameter and the 
corresponding effective critical temperature for a 
bundle consisting of a mixture of SWCNTs of these two model types.
From an experimentalist's point of view it is even more important to solve the inverse problem: 
for a given fraction $a$  of semiconductor SWCNT in the bundle and 
the experimentally determined critical  temperature $T_c(a)$,
to estimate the critical temperature 
for a bundle, consisting only of semiconductor SWCNTs $T_c(a=1)$? It will be also interesting to know, 
can one obtain $T_c$ 
much higher than in other carbon based nanostructures, and particularly higher than in alkali metal doped fullernes.

Spatial variations of the superconducting order parameter are significant for nanoscale systems, 
including nanotubes \cite{crespi,shanenko}.   
In this work we use a microscopic 
theory based on inhomogeneous Bogoliubov-de Gennes equations to establish how the superconducting properties of 
a bundle depend 
on the fraction of doped semiconductor nanotubes, with a higher SC 
order parameter.
We assume that the nanotubes in the bundle are approximately of the same radii and tightly
packed making a triangular lattice in the bundle's transverse section, with the primitive vectors $\vec{a}_1=R\vec{x}$,
$\vec{a}_2=R\vec{x}/2+\sqrt{3}/2 R \vec{y}$. Here $\vec{x},\vec{y}$ are the unit basis vectors, 
and $R$ is the intertube distance. 
The lattice can enumerated by  indexes $(i,j)$, which correspond to the position of a nanotube
 $R_{i,j}=\vec{a}_1 i+\vec{a}_2 j$, but in this work we prefer to enumerate nanotubes in a
 $N\times N$ bundle using a single index through the mapping 
$k=i+j N$, where $N$ is the number of nanotubes in a raw. 
%We assumed translation invariance along the longitude $z$ direction of the nanotubes.
In the bundle semiconductor nanotubes are assumed to occupy the fraction $a$ of the sites,
 and metallic nanotubes $1-a$, accordingly.  
 In our model the conduction electrons can freely travel along the nanotubes, and in this picture 
it corresponds for an electron staying at the same lattice site. 
However, electrons can also hop
to the neighboring  nanotubes (sites). In principle, there may be three different hoping constants,   
with the hopping matrix elements $t^{kk'}$ equal to either
$t_{mm}$,$t_{ms}$,$t_{ss}$, corresponding to the hopping between metallic-metallic (mm), 
metallic-semiconducting (ms), or semiconducting-semiconducting (ss) nanotubes. Moreover, these parameters 
may significantly fluctuate from one site to another, due to mismatch between SWCNTs of 
different chirality. 
 In the superconducting regime Cooper pairs can be formed and can freely move along the nanotubes, and can 
also hope from one tube to another.

For the description of the system  we utilized a tight-binding Hubbard Hamiltonian of the form:
\begin{eqnarray} \label{hubbard}
H_0 = \sum_{<{\bf r}_i,{\bf r}_j>, \sigma}
t^{i,j}{c^{\dagger}_{{\bf r}_i,\sigma} c_{{\bf
r}_j,\sigma}}- \mu \sum_{<{\bf r}_i>, \sigma}
{c^{\dagger}_{{\bf r}_i,\sigma} c_{{\bf
r}_i,\sigma}}+\nonumber\\\sum_{<{\bf r}_i>} U_{int}^{i}({\bf r}_i) n_\downarrow({\bf
r}_{i})n_\uparrow({\bf
r}_{i})\nonumber\\+ \sum_{<{\bf r}_i,{\bf r}_j>, \sigma,\sigma'} V_{int}^{ij} n_\sigma({\bf r}_i)
n_{\sigma'}({\bf r}_j),
\end{eqnarray} where a
quantum-mechanical operator $c^{\dagger}_{{\bf r}_i,\sigma}$ creates
an electron on site $i$ (using single indexing), the operator $c_{{\bf r}_j,\sigma}$
eliminates an electron from the site $j$, and $n_\sigma({\bf r}_i) =
c^{\dagger}_{{\bf r}_i,\sigma}c_{{\bf r}_i,\sigma}$
represents the electron density on site $i$ with the spin polarization $\sigma$. The
electron spin, $\sigma$, can point up or down. 
$U_{int}^{i}$ is on site interaction
potential. This term in a case of attractive interaction $U_{int}^i<0$ may lead to pairing in the nanotube $i$.
  $V_{int}^{ij}$ is a strength of the coupling between electrons localized at neighboring tubes $i$ and $j$.
 
Using the Boguliubov transformation, which diagonalizes the Hamiltonian Eq.(\ref{hubbard}), we arrive to 
 inhomogeneous Bogoliubov-de Gennes equations 
for the quasiparticle amplitudes on the lattice $i$
sites $(u_n({\bf r}_i),v_n({\bf r}_i))$  \cite{bdg_inhom}:
\begin{eqnarray}
\label{matrix_equation} \left(\begin{array}{cc} \hat{\xi} & \hat{\Delta}\\
\hat{\Delta}^* &-\hat{\xi}^*
\end{array}\right)\left(\begin{array}{c} u_n({\bf r}_i)\\ v_n({\bf r}_i)
\end{array}\right)=E_n\left(\begin{array}{c} u_n({\bf r}_i)\\ v_n({\bf
r}_i)
\end{array}\right),
\end{eqnarray}
where the kinetic operator $\hat{\xi}$ and the superconducting order
parameter $\hat{\Delta}$ can be represented as:
\begin{eqnarray}
\label{3} \hat{\xi}u_n({\bf r}_i)&=&- \sum_{{\boldsymbol\delta}} {t^{i,j}}
u_{n}({\bf r}_i+{\boldsymbol\delta}) +(V^s({\bf r}_i) -\mu)
u_n({\bf r}_i),\nonumber\\
{\hat\Delta v_{n}({\bf r}_i)}&=& \sum_{{\boldsymbol\delta}}
\Delta_{{\boldsymbol\delta}}({\bf r}_i)v_{n}({\bf
r}_i+{\boldsymbol\delta})+\Delta_{{\boldsymbol s}}({\bf
r}_i)v_{n}({\bf
r}_i),
\end{eqnarray}
where ${\boldsymbol\delta}$ are the
nearest neighbor vectors for a triangular lattice,  $V^s({\bf r}_i)$ is the mean-field (Hartree) potential, 
$\mu$ is the chemical potential.
$\hat{\Delta}_{{\boldsymbol s}}$ is the conventional, s-type order parameter.
One should solve
Eq.(\ref{matrix_equation})  together with the self-consistency
conditions:
\begin{eqnarray}
\label{4} \Delta_{\boldsymbol\delta}({\bf
r}_i)=\sum_n \frac{V^{ij}_{int}}{2}(u_n({\bf
r}_i+{\boldsymbol\delta})v^*_n({\bf r}_i)+\nonumber\\u_n({\bf
r}_i)v^*_n({\bf r}_i+{\boldsymbol\delta}))\tanh(E_n/2k_BT),
\end{eqnarray}
where the pairing strength $V_{int}^{ij}$ may depend on the type of CN at sites $i$ and $j$.
The s-type order parameter (within a given nanotube $i$) is simply
\begin{eqnarray}
\label{5} \Delta_{\boldsymbol s}({\bf
r}_i)=\sum_n \frac{U^{i}_{int}}{2}(u_n({\bf
r}_i)v^*_n({\bf r}_i)+\nonumber\\u_n({\bf
r}_i)v^*_n({\bf r}_i))\tanh(E_n/2k_BT),
\end{eqnarray}
Note, the summation in Eqs.(\ref{4},\ref{5})  is over the positive eigenvalues $E_n$ only.

%Note that summation also includes the continuously quantized states along the $z$ axis.
%It can be approximated by considering finite, but relatively long (up to 1000 nm nanotubes), 
%and replacement of the integral by the summation.

Here we adopted a simplified picture assuming the same constant hopping parameter $t$ between any type 
of nanotubes.  
In this work we considered for simplicity that the pairing may happen between electrons  in the same nanotube, therefore 
neglecting much weaker 
pairing mechanism between neighboring nanotubes. 
In principle,  a weak attraction mechanism may stimulate the formation of 
a Cooper pair with  one electron in one nanotube, and the second electron in one of its nearest neighbors. 
This may result in {\it co-existing order parameters} in the system, one of the order parameters with the conventional
 (s-type) symmetry, and another with unusual symmetry.
The co-existence of order parameters with different symmetries, was studied, for example, for Uranium-based 
superconducting materials \cite{uranium}.
The possibility of such pairing  on a triangular
 lattice may result in unconventional 
superconducting properties. For example, a 2D triangular lattice was
 recently considered as a test-bed for a possibility of f-wave
 spin-triplet superconductivity \cite{fwave}.

The amplitudes $u_n(\br_i), v_n(\br_i)$
obey the constraints $\int d \br (|u_{n}(\br_i)|^2+
|v_{n}(\br_i)|^2)
= 1$ for any $n$ (normalization) and $\sum_n ( |u_{n}(\br_i)|^2+
|v_{n}(\br_i)|^2)
= 1$ for any $i$, $i$ being the site index of 
the triangular lattice.

\begin{figure}[h]
\includegraphics[width=7.cm]{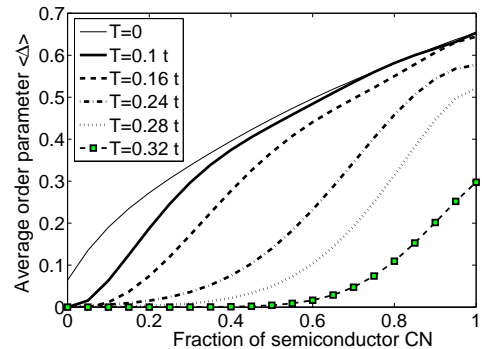}
\caption{\label{fig1} Spacially averaged superconducting order parameter  $<\Delta>$ (in units of $t$) 
as a function of fraction of semiconductor nanotubes $a$ for different temperatures.}
\end{figure} 

\begin{figure}[h]
\includegraphics[width=7.cm]{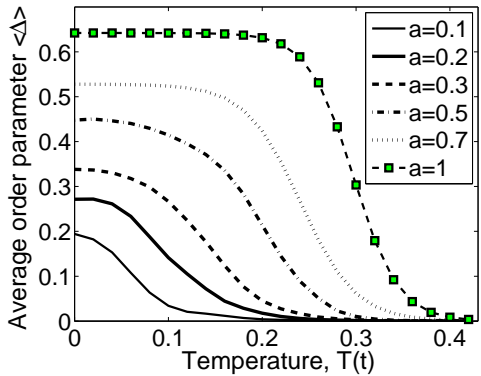}
\caption{\label{fig2} Spacially averaged superconducting order parameter  $<\Delta>$ (in units of $t$) as a function of temperature for different values 
of $a$ (fraction of semiconductor CN).}
\end{figure} 

\begin{figure}[h]
\includegraphics[width=6.cm]{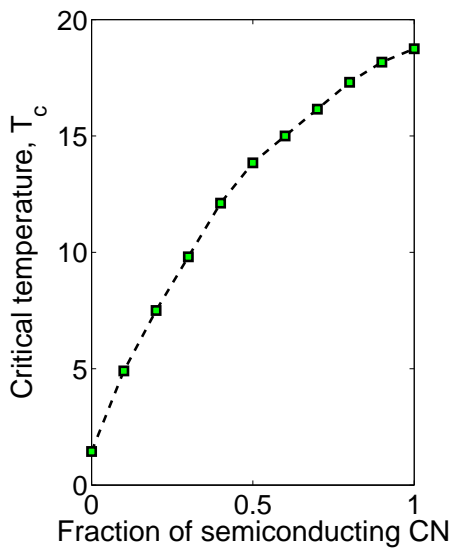}
\caption{\label{fig3} Critical temperature $T_c$ as a function of the fraction of semiconductor nanotubes $a$. 
Note a steeper slope for $a<0.5$ 
(the percolation limit in two dimensions on the triangular lattice).}
\end{figure} 

\begin{figure}[h]
\includegraphics[width=8.cm]{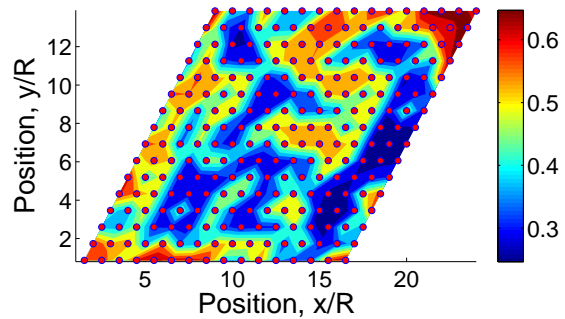}
\caption{\label{fig4} Spatial distribution of the superconducting order parameter (in units of $t$) 
at the percolation regime ($a=0.5$) and zero temperature. Red dots mark the triangular lattice.}
\end{figure} 
%\subsection{Averaged order parameter as a fraction of the metallic nanotubes in the bundle}

We studied how a spatially averaged superconducting order parameter $<\Delta>=\frac{1}{N^2}\sum_k \Delta_{\boldsymbol s}({\bf
r}_k)$ depends on the fraction of 
semiconductor SWCNTs $a$ in a $N\times N$
 bundle at different temperatures. For this purpose we generated  $P=50$ realizations for a given number of 
randomly placed  semiconductor nanotubes in the bundle.
The rest of nanotubes in the bundle are assumed to be metallic. 
To model different superconducting pairing strength for different types of nanotubes we set $U^{i}_{int}=2t$ for 
semiconducting nanotubes and $U^{i}_{int}=0.68 t$ for metallic ones, and assumed $\mu=0$ (half filled band).
In our simulations we considered $16\times16$ nunotubes in the bundle, forming a triangular lattice.

The results of calculations are shown in Fig.1. 
 At $T=0$ the order parameter
scales approximately as a square root of the fraction of the semiconducting SWCNTs in 
the bundle $T_c\propto(T^m_c-T^s_c)\sqrt{a}+T^s_c$.
 Note the convexity 
of the dependence. However,
at finite temperatures, which are between the critical temperature of a pure metallic $T^m_c$ and pure semiconducting 
$T^s_c$ bundles,  the averaged order parameter vanishes much faster with the decreasing of $a$. 
For example, at $T=0.32 t$, which is close to the critical temperature of a pure semiconducting 
SWCNT bundle $T^s_c\approx0.35 t$, the order parameter decreases in the exponential fashion with the decreasing of $a$, and 
almost vanishes at $a\approx0.5$.
%Note, also that $a=0.5$ is the known percolation threshold for a 2D triangular lattice.

In Fig.2  we also plot $<\Delta>$ as a function of temperature for several
values of $a$. One can clearly see how the order parameter vanishes above the critical temperature.
Note that with the lowering of the fraction of semiconducting nanotubes $a$, the temperature dependence 
of the order parameter shows less pronounced phase transition because of the ``dirty'' nature of inhomogeneous 
spatial distribution of the pairing properties, similar to dirty superconducting transition in case 
of large concentration of impurities and in alloys.

Using the data plotted in Fig. 2, we calculated 
how the critical temperature $T_c$ depends on the concentration of semiconducting SWCNT. 
We used $t=4.8\times 10^{-3}$eV to fit the data in [22-23], so $a=0.6$ will correspond to $T_c\approx15$K.
%and how it depends on the averaged order parameter at zero temperature
% $<\Delta(0)>$. 
In Figure 3 one can see that the critical temperature  
decreases for $a<0.5$ with a steeper slope, where $a=0.5$ corresponds to the percolation threshold in 2D triangular
 lattices (please see Fig. 4). For $a<0.5$ weaker  superconducting metallic CN are arranged in bigger size islands.
If one would take into account the phase fluctuations of the order parameter, 
this decrease of critical temperature would be even steeper, because of enhanced phase slips in relatively well isolated
semiconducting nanotubes.
 
It should be noted that our model has general applicability to any system in which there are 
two types of nanotubes, (or very thin 
nanowires) with different superconducting pairing strength are coupled in bundles. So it also describes the most
 common case of undoped pristine SWCNT bundles, which contain 30-40 $\%$ of metallic SWCNT and the rest are non-doped 
semiconducting SWCNTs, which usually do not have carriers. As has been shown this bundles have typical 
$T_c$ of 0.55 K [20-21],
 which according to our model is suppressed by the inverse proximity effect from non-superconducting 
undoped semiconducting tubes. According to our model if 100 $\%$ 
separated only metallic tubes are in the bundles, then the gap and $T_c$ should be significantly higher and we expect 
that without fluctuations accounted it can be around $T_c\approx1.3$K. 
%(Ilya. Let us estimate this from the Fig.3. I am surprised that the dependence of Tc on content is so weak 
%By the way, Fig. 3 should be changed, rescaling the Y-axis  taking into account that Tc for $60\%$ 
%fraction of Boron doped semiconducting nanotubes is 15 K (from Haruyama’s PRB paper).
Similarly for optimally doped $100\%$ semiconducting SWCNTs 
the $T_c$ should increase from observed in [22-23] $T_c\approx15$ K to
 the unsuppressed (by the inverse proximity effect of low $T_c$ metallic tubes) $T_c$ of 19-20 K.
The effect of $T_c$ suppression similar to discussed here has been observed in alkali 
metal fulelride molecular alloys of $A_x(C_{60})_x (C_{70})_{1-x}$ (24) and adding non-superconducting component, 
i.e. $C_{70}$ molecules, which do not show any superconducting pairing due to symmetry reasons 
and probably due to weaker electron-phonon coupling, strongly suppressed $T_c$ from 19 K in $100\%$ 
$C_{60}$, i.e. in $K_3C_{60}$ to $T_c$=10 K in $20 \%$
 substituted $C_{70}$ alloy. 
The experiments with selectively separated metallic and semiconducting SWCNTs, which now become available 
by new methods of effective separation 
will allow to check the validity of presented here simple model 
and to clarify the role of quantum fluctuations, 
which has not been accounted here.

We introduced  a microscopic model of superconductivity in a bundle of a mixture of carbon nanotubes.
We have studied  the dependence of a spatially averaged superconducting gap $<\Delta>$ 
 on the fraction of semiconducting
 SWCNT (having higher pairing strength) in the bundle
at different temperatures.
Note that for inhomogeneous nanoscale systems the dependence $T_c(<\Delta(T=0)>)$ 
for different concentration $a$ may be nonlinear, as a manifestation of the breakdown of the BCS theory for bulk materials.
Indeed, our calculations of $T_c(<\Delta(T=0)>)$ show a kink at $a=0.5$.
The reason is that the bundle is a highly inhomogeneous system. 
At $a<0.5$, below the percolation threshold for a 2D triangular lattice,
the bundle can be seen as a collection of finite islands of ``good'' superconductors (doped semiconducting nanotubes),
 diluted by normally conducting material (metallic nanotubes). 
Such islands demonstrate significantly suppressed superconductivity, even in the mean field description, due to the enhanced
inverse proximity effect.
 Note, that our mean-field BdG model is unable to predict and properly describe  quantum phase fluctuations of 
the order parameter in quasi one dimensional systems, 
where the superconductivity will be suppressed even stronger. 
Future research using, for example, Ginzburg-Landau inhomogeneous equations 
\cite{vinokur,qf,phase1,wire_quantum_fluct} is necessary to describe such kind of effects. 
Because the dynamics of Cooper pairs in doped carbon nanotubes can be 
more close to the diffusive regime, the Usadel equations can be applied to calculate the finite conductivity
 at $T<<T_c$ \cite{usadel}.

%The transport properties in the bundles of carbon nanotubes could be possibly described in terms of the ``freeway model''\cite{freeway}.
%PLOT LOCAL DOS!
%\end{multicols}

\end{document}